\documentclass[prl,aps,amsmath,amssymb,superscriptaddress,twocolumn]{revtex4}

\usepackage{graphicx}

\newcommand{\nc}{\newcommand}
\nc{\beq}{\begin{equation}}
\nc{\eeq}{\end{equation}}
\nc{\beqa}{\begin{eqnarray}}
\nc{\eeqa}{\end{eqnarray}}

\def\gsim{\mathrel{\rlap{\lower4pt\hbox{\hskip1pt$\sim$}}
    \raise1pt\hbox{$>$}}}       

\begin{document}

\title{Everything is Entangled}

\author{Roman~V.~Buniy} \email{roman.buniy@gmail.com}
\affiliation{Schmid College of Science, \\
Chapman University, Orange, CA 92866}

\author{Stephen~D.~H.~Hsu} \email{hsu@uoregon.edu}
\affiliation{Institute of Theoretical Science \\ University of Oregon,
Eugene, OR 97403 \\  }

\begin{abstract}
We show that big bang cosmology implies a high degree of entanglement of particles in the universe. In fact, a typical particle is entangled with many particles far outside our horizon. However, the entanglement is spread nearly uniformly so that two randomly chosen particles are unlikely to be {\it directly} entangled with each other -- the reduced density matrix describing any pair is likely to be separable.
\end{abstract}


\maketitle

\date{today}

\bigskip

\section{Ergodicity and properties of typical pure states}

When two particles interact, their quantum states generally become entangled. Further interaction with other particles spreads the entanglement far and wide. Subsequent local manipulations of separated particles cannot, in the absence of quantum communication, undo the entanglement. We know from big bang cosmology that our universe was in thermal equilibrium at early times, and we believe, due to the uniformity of the cosmic microwave background, that regions which today are out of causal contact were once in equilibrium with each other. Below we show that these simple observations allow us to characterize many aspects of cosmological entanglement.

We will utilize the properties of {\it typical} pure states in quantum mechanics. These are states which dominate the Hilbert measure. The ergodic theorem proved by von Neumann \cite{vN} implies that under Schrodinger evolution most systems spend almost all their time in typical states. Indeed, systems in thermal equilibrium have nearly maximal entropy and hence must be typical. Typical states are maximally entangled (see below) and the approach to equilibrium can be thought of in terms of the spread of entanglement.

\begin{figure}
\includegraphics[width=8cm]{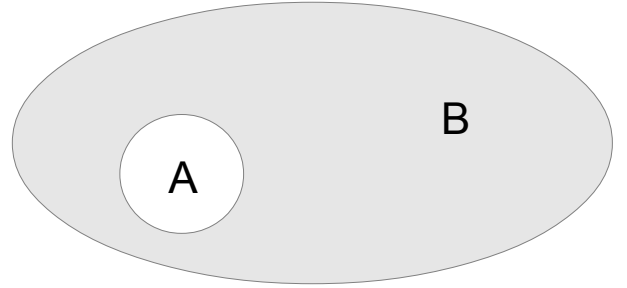}
\caption{The entire system is in a pure state $\psi$ subject to a constraint on total energy. Tracing over the shaded region B yields a density matrix $\rho_A$. For typical $\psi$ (which dominate the set of possible states), $\rho_A$ is nearly thermal and nearly maximally entangled.}
\label{AB}
\end{figure}

Consider a large system subject to a linear constraint $R$ (for example, that it be in a superposition of energy eigenstates with the energy eigenvalues all being near some $E_\ast$), which reduces its Hilbert space from ${\cal H}$ to a subspace ${\cal H}_R$. Divide the system into a subsystem $A$, to be measured, and the remaining degrees of freedom which constitute an environment $B$, so ${\cal H} = {\cal H}_A \otimes {\cal H}_B$ and 
 \begin{equation}
 \rho_A \equiv \rho_A(\psi)= {\rm Tr}_B \vert \psi \rangle \langle \psi \vert
\end{equation}
is the density matrix which governs measurements on $A$ for a given pure state $\psi$ of the whole system. Note the assumption that these measurements are local to $A$, hence the trace over $B$. (See Fig. \ref{AB}.)

It can be shown \cite{Winters1,intuition} (see also \cite{Gemmer,Goldstein}), using the concentration of measure on hyperspheres \cite{HDG} (Levy's Lemma), that for almost all $\psi \in {\cal H}_R$, 
\begin{equation}
\rho_A(\psi) \approx {\rm Tr}_B\left(\rho_*\right) \equiv \Omega_A\,,
\end{equation}
where $\rho_* = { \bf 1}_R / d_R$ is the equiprobable maximally mixed state on the restricted Hilbert space ${\cal H}_R$ (${ \bf 1}_R$ is the identity projection on ${\cal H}_R$ and $d_R$ the dimensionality of ${\cal H}_R$). $\Omega_A = {\rm Tr}_B \left(\rho_*\right)$ is the corresponding canonical state of the subsystem $A$. The result holds as long as $d_B \gg d_A$, where $d_{A}$ and $d_{B}$ are the dimensionalities of  the ${\cal H}_A$ and ${\cal H}_B$ Hilbert spaces. (Recall that these dimensionalities grow exponentially with the number of degrees of freedom. The Hilbert space of an $n$ qubit system is $2^n$--dimensional.) In the case of an energy constraint $R$, $\Omega_A$ describes a perfectly thermalized subsystem with temperature determined by the total energy of the system (i.e., a micro canonical ensemble).

To state the theorem in \cite{Winters1} more precisely, the (measurement-theoretic) notion of the \emph{trace-norm} is required, which can be used to characterize the distance between two mixed states $\rho_A$ and $\Omega_A$:
\begin{equation}
\Vert\rho_A-\Omega_A\Vert_1\equiv{\rm Tr}\sqrt{\left(\rho_A-\Omega_A\right)^2}\,.
\end{equation}
This quantifies how easily the two states can be distinguished by measurements, according to the identity
\begin{equation}
\label{tracesup}
\Vert\rho_A-\Omega_A\Vert_1 = {\rm sup}_{\Vert O\Vert\leq1}\,{\rm Tr}\left(\rho_AO-\Omega_AO\right)\,,
\end{equation}
where the supremum runs over all observables $O$ with operator norm $\Vert O\Vert \leq$ 1. The trace on the right-hand side of (\ref{tracesup}) is the difference of the observable averages $\langle O\rangle$ evaluated on the two states $\rho_A$ and $\Omega_A$, and therefore specifies the experimental accuracy necessary to distinguish these states in measurements of $O$.

The theorem then states that (for $\epsilon > 0$)
\begin{align}
\label{w1}
& {\rm Prob} \left[
\Vert\rho_A\left(\psi\right)-{\rm Tr}_B\left(\rho_*\right)\Vert_1 ~\geq~ \epsilon+ d_A  d_R^{\, -1/2}
\right]  \nonumber \\
&~~~~~~~~~~~~~~~~~ <~  2\exp ( -\epsilon^2d_R/18\pi^3 )~.
\end{align}
In words: let $\psi$ be chosen randomly (according to the Haar measure on the Hilbert space) out of the space of allowed states ${\cal H}_R$; 
the probability that a measurement on the subsystem $A$ \emph{only}, with measurement accuracy given by the rhs of (\ref{w1}), will be able to tell the pure state $\psi$ (of the entire system) apart from the maximally mixed state $\rho_*$ is exponentially small in $d_R$, the dimension of the space ${\cal H}_R$ of allowed states. Conversely, for almost all pure states $\psi$ any small subsystem $A$ will be found to be extremely close to perfectly thermalized (assuming the constraint $R$ on the whole system was an energy constraint).

As mentioned, the overwhelming dominance of typical states $\psi$ is due to the geometry of high-dimensional Hilbert space and the resulting concentration of measure. It is a consequence of kinematics only -- no assumptions have been made about the dynamics. Almost any dynamics -- i.e., choice of Hamiltonian and resulting unitary evolution of $\psi$ -- leads to the system spending nearly all of its time in typical states for which the density matrix describing any small subsystem $A$ is nearly thermal \cite{vN,Winters2}. Typical states $\psi$ are maximally entangled, and the approach to equilibrium can be thought of in terms of the spread of entanglement, as opposed to the more familiar non-equilibrium kinetic equations. 

Since generic pure states tend to evolve into typical states, any mixture of pure states is likely to evolve into a mixture of typical states. Hence, our analysis does not require any specific assumptions about whether the system (i.e., the universe) is in a pure or mixed state. If it is in a mixture, we simply have (classical) probabilities of finding the system in one of two or more typical pure states. For simplicity, in the rest of the paper we will always assume the system as a whole is in a pure state.

We can restate these results in terms of the entanglement entropy of the subsystem $A$, thereby making contact with the Second Law of Thermodynamics. The entanglement entropy is simply the von Neumann entropy of $\rho_A$:
\begin{equation}
\label{entanglemententropy}
S (\rho_A) = - {\rm Tr} \, \rho_A \ln \rho_A\,.
\end{equation}
Using the same results on the concentration of measure, it can be shown \cite{Hayden} that, for the overwhelming majority of pure states $\psi$, $S( \rho_A )$ is extremely close to its maximum value $\ln d_A$:
\begin{align}
& {\rm Prob} \left[ S( \rho_A) < \ln d_A - \alpha - \beta \right] \nonumber \\
& ~~~~~~ \leq \exp \left[ -  \frac{(d_A d_B - 1) C \alpha^2}{(\ln d_A)^2} \right] \, ,
\end{align}
where $\alpha > 0$, $\beta = d_A / d_B$ and $C = (8 \pi^2)^{-1}$. This implies \cite{vN,Winters2} that, for almost any choice of dynamics, a subsystem $A$ is overwhelmingly likely to be found with nearly maximal entropy $S( \rho_A )$. The Second Law is seen to hold, in a probabilistic sense, even though the underlying dynamics is time-reversal invariant: density matrices $\rho_A$ with small entropy are highly improbable, and if $A$ is found in a low-entropy state, the entropy is more likely to increase than decrease over any macroscopic time interval. 

\section{Cosmology}

In the following discussion we assume a semiclassical space-time and focus on field-theoretic degrees of freedom (e.g., particles such as photons or electrons). Although the analysis takes place in curved space, quantum gravitational effects are never significant, and the rules of ordinary quantum mechanics apply throughout (just as they do in the Earth's gravitational field). We adopt a cosmological time coordinate (e.g., that of the FRW metric) and evolve the collective wave function of particles using the Schrodinger equation in those coordinates. 

The cosmic microwave background provides direct evidence that the universe was in thermal equilibrium at temperatures of order eV. Nucleosynthesis of light elements provides indirect evidence of thermal equilibrium at temperatures in the keV to MeV range. This suggests that the state describing the universe in the past was typical. The ergodic theorem \cite{vN}, or equivalently, the Second Law, implies that the universe is likely to be in a typical state today. Thus the entanglement entropy of any subsystem $A$ is likely close to maximal.

To proceed further we recall that a cosmological horizon volume is the largest region over which causal contact is possible. The size of this region is
\begin{equation}
\label{horizon}
d_H (t) = a(t) \int_{0}^t \frac{dt'}{a(t')}~~,
\end{equation}
where $a(t)$ is the FRW scale factor. Our present horizon volume consists of many sub-regions that are only now coming into causal contact, at least as implied by (\ref{horizon}) under ordinary (e.g., radiation- or matter-dominated) expansion. The fact that the temperature and distribution of density perturbations (not to mention stars and galaxies) are approximately uniform over these regions suggests that, somehow, they were already in causal contact during some previous epoch. Most researchers now believe that this is due to a period of exponential growth in $a(t)$ known as inflation. During this era the metric was approximately that of de Sitter space and the energy density was dominated by the vacuum energy of the inflaton field. In this scenario, the currently visible universe originated from a progenitor region much smaller than the horizon volume at the start of inflation: 
\begin{equation}
\label{visible}
r_0 \ll d_H (t_{\rm inflation} )~~.
\end{equation} 
Because the entire horizon volume at $t = t_{\rm inflation}$ was in equilibrium, all of our visible universe and regions which will only become visible in billions of years experienced similar initial conditions, thus explaining the observed homogeneity and isotropy. At the quantum level, this equilibrium assumption implies that the pure state $\psi$ describing a region of size $d_H (t_{\rm inflation} )$ at the start of inflation was typical. Due to the inequality (\ref{visible}), entanglement today must extend far beyond the currently visible universe. In fact, as we show below, particles in our horizon volume are mostly entangled with particles outside of it.

We can see explicitly how entanglement is transferred by considering the inflaton field in the standard model of slow-roll inflation. Before inflation begins, the inflaton and other degrees of freedom are in thermal equilibrium and we expect their states to be typical. Once the inflaton vacuum energy begins to dominate the stress energy tensor, the universe supercools and the gravitational dynamics is determined by the semiclassical evolution of the scalar field as it slowly rolls along its nearly flat potential. However, at the quantum level the inflaton wave function at each position in space is still entangled with the wave function at other positions: due to the non-zero de Sitter temperature and inflaton-graviton scattering, there are interactions which ``measure'' the value of the inflaton field and entangle its wave function with nearby degrees of freedom. (Indeed, this has to be the case for a semiclassical space-time to emerge, whose dynamics is mainly driven by the vacuum energy of the inflaton.) We need only require that the ergodic theorem apply to the system comprised of inflaton field, background particles and gravitons, during an epoch in which gravitational effects are small (e.g., the de Sitter timescale is larger than particle interaction timescales). Under this assumption, the wave functions of the inflaton field in different patches of each horizon volume are entangled. 

Due to the exponential expansion in de Sitter space the physical separation between points grows superluminally and regions which were originally in the same horizon volume of size $d_H (t_{\rm inflation} )$ become space-like separated. Local evolution in each causally separate region cannot undo the pre-existing entanglement. When the inflaton field decays, causing particle production and reheating the universe, this entanglement is transferred to the decay products, which include the particles that make up the universe today. The discussion above is in the context of a specific model of inflation, but in general we expect any mechanism for superluminal expansion which solves the isotropy and homogeneity problems will lead to entanglement across many horizon volumes.

To summarize, modern cosmology suggests that most of the particles in the visible universe exhibit a high degree of entanglement with degrees of freedom far beyond our horizon volume. While it is true that gravitational clumping (e.g., of galaxies or stars) \cite{gravity} allows local deviation from thermal equilibrium, entanglement with causally separated regions produced in earlier cosmological epochs cannot be removed by subsequent local dynamics.

\bigskip

\section{Aspects of maximal entanglement}

I. {\it Maximal entanglement and Schmidt decomposition}

We argued above that any small subsystem $A$ (``small'' here includes our entire horizon volume today!) is maximally entangled with the rest of the universe (most of which is not yet visible to us). That is, $S (A) \approx \ln d_A$. We can further interpret this using the Schmidt decomposition theorem: for any 
pure state $\psi_{AB}$ of a composite system $AB$, there exist orthonormal states $
\psi^{(n)}_A$ for system $A$ and $\psi^{(n)}_B$
for system $B$ such that
\begin{equation}
\label{SD}
\psi_{AB}  = \sum_n ~\lambda_n^{1/2} ~ 
\psi^{(n)}_A \,  \psi^{(n)}_B  \, ,
\end{equation}
where $\lambda_n^{1/2}$ are nonnegative real numbers satisfying
$\sum_n \lambda_n = 1$. This is a simple consequence of the singular
value decomposition theorem. The dimensions of ${\cal
H}_A$ and ${\cal H}_B$ can be very different, and the range
over which the sum in Eq.~(\ref{SD}) runs is determined by the smaller
Hilbert space. Note that the Schmidt decomposition states might be quite complex -- possibly involving superpositions of many degrees of freedom. Tracing over $B$ yields a density matrix $\rho_A$ with eigenvalues $\lambda_n$. From the previous discussion we know that all $\lambda_n \approx 1 / d_A$.

A measurement of subsystem $A$ which determines it to be in state $\psi^{(n)}_A$ implies that the rest of the universe must be in state $\psi^{(n)}_B$. For example, $A$ might consist of a few spins \cite{FN1}; it is interesting, and perhaps unexpected, that a measurement of these spins places the rest of the universe into a particular state $\psi^{(n)}_B$. As we will see below, in the cosmological context these modes are spread throughout the universe, mostly beyond our horizon. Because we do not have access to these modes, they do not necessarily prevent us from detecting $A$ in a superposition of two or more of the $\psi^{(n)}_A$. 

However, if we had sufficient access to $B$ degrees of freedom (for example, if the relevant information differentiating between $\psi^{(n)}_B$ states is readily accessible in our local environment or in our memory records), then the $A$ system would decohere into one of the $\psi^{(n)}_A$.

\bigskip

\begin{figure}
\includegraphics[width=8cm]{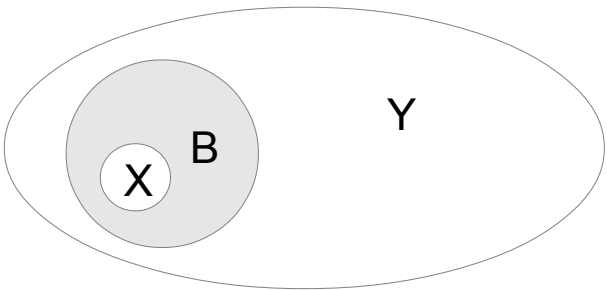}
\caption{Tracing over the shaded region $B$ yields a density matrix $\rho_{XY}$. If the dimensionality $d_B$ is much smaller than $d_X d_Y$, the entanglement of formation $E_f ( \rho_{XY} )$ will be nearly maximal. Taking $B$ to be a horizon volume, this implies that $X$ is mostly entangled with degrees of freedom outside its own horizon.}
\label{XBY}
\end{figure}

II. {\it Entanglement across horizons}

Consider Fig. (\ref{XBY}), where $B \cup X$ is our horizon volume, with $X$ a small subregion. Tracing over $B$ yields a density matrix $\rho_{XY}$ describing the entanglement of region $X$ with the rest of the universe $Y$ (all of which is outside the currently visible universe; $Y$ is assumed much larger than $B$). Because entanglement should be roughly uniformly distributed over degrees of freedom in a typical state, we expect that most of the entanglement entropy $S( \rho_{XY} )$ (which must be nearly maximal) is with modes in $Y$ rather than $B$. Indeed, one can show (theorem V.1 of \cite{Hayden}) that if $d_B \ll d_X d_Y$ (i.e., many more degrees of freedom in $X \cup Y$ than in $B$), it is exponentially likely that the {\it entanglement of formation} $E_f (XY)$ is close to $\log ( d_X d_Y )$ (i.e., is also nearly maximal). The entanglement of formation is a measure of entanglement for mixed states, such as $\rho_{XY}$ \cite{Bennett}. It is defined as the minimum entanglement resource necessary to create $\rho_{XY}$ without further quantum communication. Alternatively, $E_f (XY)$ is equal to the least expected entanglement of any ensemble of pure states which realize $\rho_{XY}$. That is, for all decompositions
\begin{equation}
\rho_{XY}  =  \sum \, w_i \,  \vert \phi^i_{XY} \rangle \langle \phi^i_{XY} \vert~~,
\end{equation}
where $w_i$ are probabilities and $\phi^i_{XY}$ is a pure state, $E_f (XY)$ is the minimum expected entanglement $\sum w_i S( \phi^i_{XY} )$. These statements imply that most of the entanglement entropy $S( \rho_{XY} )$ is due to entanglement with modes of $Y$, which are causally disconnected (space-like separated) from $X$.

\bigskip

III. {\it Small systems are likely to be in separable states}

\begin{figure}
\includegraphics[width=8cm]{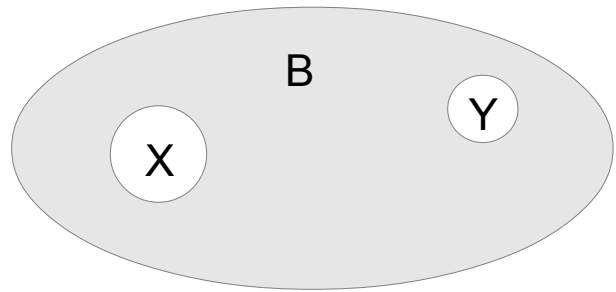}
\caption{Tracing over the shaded region $B$ yields a density matrix $\rho_{XY}$. If $X$ and $Y$ are small subsystems ($(d_X d_Y)^2 \ll d_B$), $\rho_{XY}$ is likely to be separable -- $X$ and $Y$ are not directly entangled with each other. Note $X$ and $Y$ need not be widely separated; the figure is drawn this way for visual clarity.}
\label{XYB}
\end{figure}

Fig. (\ref{XYB}) depicts two small regions $X$ and $Y$ (although depicted as far apart, they could also be spatially proximate). For example, each could consist of a single or small number of individual particles. The approximately uniform distribution of entanglement over all degrees of freedom in a typical state suggests that $X$ and $Y$ share only negligible entanglement {\it directly} with each other. A measure of this direct entanglement is again the entanglement of formation for the density matrix $\rho_{XY}:~ E_f (XY)$, which we expect to be small. Indeed, theorem V.1 in \cite{Hayden} provides an upper bound on $E_f (XY)$ which vanishes in the limit that $d_B$ is much larger than $d_X d_Y$. When $d_B \gg (d_X d_Y)^2$, $\rho_{XY}$ is exponentially likely to be separable:
\begin{equation}
\rho_{XY} = \sum w_i \, \rho^i_X \otimes \rho^i_Y~~,
\end{equation}
where $w_i$ are real, positive and sum to unity, and $\rho^i_X $ and $\rho^i_Y$ are density matrices on $X$ and $Y$. Separable states may exhibit classical correlations, but no entanglement. Even the classical correlations must be small because we know that $\rho_{XY} \approx {\bf 1}_{XY} / (d_X d_Y)$.

\section{Conclusions}

The cosmological quantum state is likely to be typical in a Hilbert space describing degrees of freedom over a region many times as large as the visible universe (our current horizon volume). This implies a high degree of entanglement of particles, with the entanglement distributed uniformly over most of the degrees of freedom. As a consequence, small subsystems are mostly entangled with particles far beyond the horizon, and two randomly chosen small subsystems are unlikely to be directly entangled with each other.

\bigskip

\emph{Acknowledgements---}  SH is supported by the Department of Energy under DE-FG02-96ER40969 and by the National Science Foundation under Grant No. NSF PHY11-25915. He thanks KITP UCSB (the Bits, Branes, Black Holes workshop) for its hospitality while this work was completed.


\bigskip



\baselineskip=1.6pt
\begin{thebibliography}{99}



\bibitem{vN} J. von Neumann, Zeitschrift fuer Physik 57: 30-70 (1929); recent English translation: arXiv:1003.2133. See also S. Goldstein, J. L. Lebowitz, R. Tumulka, N. Zanghi, European Phys. J. H 35: 173-200 (2010), arXiv:1003.2129.



\bibitem{Winters1}
 S. Popescu, A. J. Short and A. Winter, {\it Nature Physics} {\bf 2}, 754 (2006), arXiv:quant-ph/0511225.

\bibitem{intuition} We can motivate the result by considering a spin state of the form $\psi = (1 / 2^N) \sum \exp (i \theta ( \sigma_1, \dotsc ,\sigma_N ) ) \, \vert \sigma_1, \dotsc ,\sigma_N \rangle$, where $\sigma_i = \pm$. Tracing over all but the first $n$ spins ($N \gg n$) yields an approximately diagonal density matrix $\approx {\bf 1} / 2^n$. The off-diagonal entries are small for generic (random) functions $\theta$ due to cancellations. Our example is only illustrative, however, because $\psi$ is not entirely typical -- we have assumed equal probabilities for $\psi$ to be found in each $\vert \sigma_1, \dotsc , \sigma_N \rangle$ state.

\bibitem{Gemmer}
J. Gemmer, M. Michel and G. Mahler, {\it Quantum Thermodynamics: Emergence of Thermodynamic Behavior Within Composite Quantum Systems} (Springer, 2004).

\bibitem{Goldstein}
S. Goldstein, J. L. Lebowitz, R. Tumulka, and N. Zanghi,
Phys. Rev. Lett. 96, 050403 (2006).


\bibitem{HDG}
M. Ledoux, {\it The concentration of measure phenomenon} (American Mathematical Society, 2001).

\bibitem{Winters2}
N. Linden, S. Popescu, A. J. Short, A. Winter, arXiv:0812.2385.

\bibitem{Hayden}
P. Hayden, D. W. Leung and A. Winter, Comm. Math. Phys. 265, 95 (2006).

\bibitem{gravity} Whether and how the ergodic theorem applies to strong gravity is an open question. See, e.g., S.~D.~H.~Hsu and D.~Reeb,
  Mod.\ Phys.\ Lett.\ A {\bf 24}, 1875 (2009), arXiv:0908.1265.

\bibitem{FN1} It is possible that $A$ is atypical (out of equilibrium) and therefore not covered by the results stated here. However, a randomly chosen subset $A$ is overwhelmingly likely to be typical -- most of the entropy in the universe today is either in black holes or in thermal relics such as the CMB. See P.~Frampton, S.~D.~H.~Hsu, D.~Reeb and T.~W.~Kephart,
  Class.\ Quant.\ Grav.\  26, 145005 (2009),  arXiv:0801.1847.
 
\bibitem{Bennett} C. H. Bennett, D. P. DiVincenzo, J. A. Smolin, and W. K. Wootters, Phys. Rev. A, 54:3824�3851 (1996), arXiv:quant-ph/9604024.  
  
\end{thebibliography}
\end{document}